\documentclass[prb,aps]{revtex4}

\usepackage{graphicx}
\usepackage{epsfig}
\usepackage{dcolumn}
\usepackage{bm}

\begin{document}

\title{Transport via a quantum shuttle}

\author{A.D. Armour\footnote{Current address: School of Physics and
Astronomy, University of Nottingham, Nottingham NG7 2RD, United Kingdom} and A. MacKinnon}
\affiliation{The Blackett Laboratory, Imperial College of Science, Technology
 and Medicine, London SW7 2BW, United Kingdom}%

\date{\today}

\begin{abstract}
We investigate the effect of a quantised vibrational mode on
electron tunneling through a chain of three quantum dots. The
outer dots are coupled to voltage leads, but the position of the
central dot is not rigidly fixed. Motion of the central dot
modulates the size of the tunneling barriers in opposite ways so
that electron tunneling is correlated with the position of the
oscillator. We treat the electronic part of the problem using a
simple Coulomb-blockade picture, and model the vibration of the
central dot as a quantum oscillator. We calculate the
eigenspectrum of the system as a function of the energy level
shift between the outer dots. Using a density matrix method, we
include couplings to external leads and calculate the steady-state
current through the device. The current shows marked resonances
which correspond to avoided-level crossings in the eigenvalue
spectrum. When the tunneling length of the electrons is of order
the zero-point position uncertainty of the quantum oscillator,
current far from the electronic resonance is dominated by
electrons hopping on and off the central dot sequentially; the
oscillator can be regarded as shuttling electrons across the
system [L.Y. Gorelik, et al., Phys. Rev. Lett. {\bf{80}}, 4526
(1998)]. Damping of the oscillator can increase the current by
preventing electrons from hopping `backwards'.

\end{abstract}

\maketitle

\section{Introduction}

Electron tunneling rates through mesoscopic junctions are strongly modulated by
changes in the physical extent of the barriers which can arise from excitation of
vibrational degrees of freedom.\cite{jn,jn2} The tunneling rate
of electrons through insulating barriers has an exponential dependence on the
physical extent of the barriers and so is sensitive to even relatively small
changes in their size. In standard semiconductor structures the components are
rigidly fixed so that fluctuations in the widths of tunnel barriers  are relatively
unimportant. However, whenever electron transport occurs across flexible tunnel
barriers modulation of the barrier widths due to vibrations  is expected to
affect the tunneling. The perturbation of tunneling processes across a single mesoscopic
junction by a mechanical degree of freedom has been analysed in some detail, and
is now relatively well understood.\cite{jn,jn2} However, when there are two or more
flexible junctions connected in series more complicated effects arise, in particular
excitation of vibrational degrees of freedom can lead to shuttling of electrons
between the junctions.

The idea of a mesoscopic electron  shuttle in which a vibrational
degree of freedom modulates tunneling rates across two junctions
in turn was proposed recently by  Gorelik {\emph{et al.}}\cite{Gk}
(although a fully classical charge shuttle is far from
new\cite{class}).  Gorelik {\emph{et al.}} considered a system in
which electrons are transported between two leads by a small
metallic grain embedded in an elastic medium. The interaction
between charges on the grain and the  leads couples to the
position of the oscillator leading to a dynamic instability.
Electrons tunnel onto the grain from one of the leads, resulting
in a Coulombic repulsion which drives the grain towards the other
lead where the electrons tunnel off: hence vibrations of the grain
shuttle electrons between leads. In the model of  Gorelik
{\emph{et al.}} the  electron tunneling is treated within the
orthodox theory of single electron tunneling, whilst the elastic
medium is treated as a classical oscillator.

A mesoscopic shuttle device similar in some respects to that proposed by
Gorelik {\emph{et al.}} has been fabricated by Blick {\emph{et al.}}
\cite{blick,blick2}
The device closely resembles a miniaturised bell clapper, consisting
of two tunneling contacts with a cantilever in between. However, the cantilever
does not `ring' (i.e. shuttle electrons) spontaneously, but instead is driven by strong
gate voltages applied across the cantilever itself (away from the tunneling contacts).
 The rate of electron tunneling can be controlled by the amplitude of the drive
  when it is tuned to one of the natural frequencies of the
   cantilever. Electron transport
 in this system can again be understood within orthodox
  single electron tunneling theory, modified to account for position dependent
  behavior in the tunneling rates.\cite{blick2,weiss}

Systems in which electron transport between two contacts is
mediated by a vibrational mode of a self-assembled structure have
also been investigated.\cite{nat,sa} The most striking example of
such a system is the $C_{60}$ single electron transistor,
fabricated by Park {\emph{et al.}}\cite{nat} In this device, a
single  $C_{60}$ molecule was deposited in a narrow gap between
gold electrodes. The current flowing through the device was found
to increase sharply whenever the applied voltage was sufficient to
excite vibrations of the molecule about the minima of the van der
Waals potential between it and the electrodes, or an internal mode
of the molecule itself.\cite{nat,boese,nish2}

The device fabricated by Park
{\emph{et al.}} is an example of a molecular electronic device\cite{molec}
in which electrical conduction occurs through single molecules connected to
conventional leads. The junctions between molecular components
and leads will be much more flexible than those in conventional solid-state
nanostructures and fluctuations in their width may modify their current
characteristics significantly. Furthermore, vibrational modes
of the molecular components themselves may play an important role in
determining the transport properties.\cite{gaud}

Variations in the widths of tunnel barriers are also expected to have an
important effect on the transport properties of nanoparticle chains.\cite{nish}
Nanoparticle chains consist of small metal grains stabilised by ligands, with
electronic transport occurring via tunneling between the metal particles. Because
of the relative softness of the ligand matrix, vibrations of the metal grains
can significantly modify the electronic tunneling rates.

In the present work, electron tunneling through an
 electromechanical system in the extreme quantum mechanical and
 Coulomb blockade limits is investigated.  Our interest is focused on the case
 where the width of the tunnel barriers for electrons are modulated by a quantum
 mechanical vibrational degree of freedom. We analyse a model shuttle system consisting of
 a row of three quantum dots in which the central dot is
mounted on a quantum harmonic oscillator.

The model considered here is simplified in many respects so that it is possible to
build up an understanding of the characteristics of the transport which
occurs when there is strong coupling between the electronic degrees of freedom
and the displacement of a quantum harmonic oscillator.
We assume that the capacitance of
the central grain or dot, and the outer dots  is sufficiently
high that transport is confined to the Coulomb blockade regime. Furthermore,
we assume that the electronic states of the system are coherent.
We use the language of semiconductor nanostructures, describing our system in
terms of quantum dots, because it provides a convenient short-hand for what is
in effect a localised electronic state. In practice, what we refer to
quantum dots could be actual semiconductor dots, large molecules or
metallic nanoparticles. Similarly, the oscillator could be an ultra-high
frequency mechanical resonator with a dot mounted on its tip, like that
studied by Blick {\emph{et al.},\cite{blick} alternatively the oscillator mode could
arise from vibration  of the central `dot' within a stabilising elastic
medium or the potential  confining it between the outer dots.

We find that the current characteristics of the model shuttle can
largely be understood by analysing the eigenspectrum of the
isolated system of three dots and the quantum oscillator. Tunnel
coupling of the dot states, to each other and to the position of
the oscillator, leads to repulsion of the eigenvalues and mixing
of the eigenstates associated with states localised on individual
dots. The mixed states consist of superpositions of the states
associated with the individual dots and hence lead to
delocalisation of the electronic states between the dots. Analysis
of the current which flows when the shuttle is weakly coupled to
leads, reveals strong resonances  corresponding to the occurrence
of the delocalised states. The current through the shuttle is
found to depend sensitively on the amount by which the oscillator
is damped, the strength of the couplings between the dots and the
background temperature. A preliminary account of some of these
findings has been given elsewhere.\cite{us}

The organisation of the paper is as follows. In section II we describe the
details of our model shuttle system. We introduce the Hamiltonian of the
system and describe how coupling to external leads can be taken into account.
In section III we examine the eigenspectrum
of the system as a function of the difference in energy between the levels in
the fixed dots. The current characteristics of the device are described in
section IV. A summary and discussion of the results is given in section V.
Details of how the Hamiltonian can be approximated by a finite matrix are
given in appendix A and the derivation of the density matrix equation of
motion is outlined in appendix B.

\section{Model Formulation}

We begin by detailing the  model Hamiltonian which we
use to describe the behavior of the shuttle system.  The shuttle consists of
a linear chain of three quantum dots and a single vibrational mode, as shown
schematically in Fig.\ 1. The physical locations
of the two outer dots are fixed, while the central dot is
mounted on the vibrational mode whose behavior is modelled by a quantum
oscillator. The  electronic part of the system is idealised:
each of the dots is represented  by a single, localised, energy state.
The dynamics of electron transport through
the system is analysed using the density matrix formalism as this allows couplings
to external leads and the effects of the oscillator's environment to be
incorporated most conveniently.

\subsection{Tight--binding model}
Within the Coulomb blockade regime, the charging energy of adding more than one
electron to the shuttle is assumed to be sufficiently high that only one
transport electron can occupy the chain of three dots at any one time.
Therefore, the electronic degrees of freedom of the
isolated system can be represented completely by the state in which none of
the dots are occupied\cite{dd1,dd1a} $|0\rangle$ and the
three localised states associated with the lefthand, central and righthand
dots: $| l\rangle$, $| c\rangle$ and $|r\rangle$ respectively. The system  is modelled by a
tight-binding Hamiltonian of the form
\begin{eqnarray}
H&=&\varepsilon_l|l\rangle\langle l|+\varepsilon_r|r\rangle\langle r|
+\varepsilon_c(\hat{x})|c\rangle\langle c|+
\hbar\omega\hat{d}^{\dagger}\hat{d} \\ \nonumber
&&-V{\mathrm{e}}^{-\alpha[x_0+\hat{x}]}
(|c\rangle\langle l|+| l\rangle\langle c|)\\ \nonumber
&&-V{\mathrm{e}}^{-\alpha [x_0-\hat{x}]}
(|c\rangle\langle r|+| r\rangle\langle c|),
\end{eqnarray}
where $| i\rangle\langle i |$, $i=l,c,r$, are projection
operators for the three  electronic states and the vibrational mode, frequency
$\omega$, is operated on by  $\hat{d}$. The tunneling
elements between the dots depend exponentially on the displacement operator of
the vibrational mode, $\hat{x}=\Delta x_{zp}(\hat{d}^{\dagger}
+\hat{d})$, where $\Delta x_{zp}$ is the zero-point position uncertainty of the
oscillator. The oscillator position has an expectation value of zero when
the oscillator is unperturbed. The tunneling amplitude and
length are given by the positive quantities $V$ and $1/\alpha$
respectively. Although we have not stated it explicitly, it is understood that
the potential energy of the vibrational modes has a hard wall cut-off at the
left and righthand dots (i.e., at $x=\pm x_0$).

The energies of the outer two dots, $\varepsilon_{l(r)}$,
together define a voltage bias across the device, $eV_b=\varepsilon_b=\varepsilon_l-\varepsilon_r$.
The energy of the central level is set by its position between the outer
dots as we assume that it undergoes a Stark shift proportional to its position\cite{three}
so that $\varepsilon_c=\varepsilon_l-(\hat{x}+x_0)\varepsilon_b/2x_0$.\cite{comment}
The energy levels in the outer dots are set by external gates whose capacitance
is assumed to be much larger than the capacitances of the other junctions
(i.e.\ the junctions between the dots and the junctions between the
outer dots and the leads), but still small enough that only one
of the dots can be occupied by an electron at any one time.

The behavior of the shuttle system is readily investiagted by numerical
methods. The Hamiltonian and density operators are represented as matrices,
as described in appendix A, and the time evolution of the system can be
obtained numerically for any initial form of the density matrix.
However, the vibrational degree of freedom is only completely specified by an
infinite set of states which must be truncated for numerical calculations. The
necessary truncation is best performed in the basis of the (unperturbed) energy
eigenstates of the vibrational mode (see appendix A).
So long as the energy of the highest energy state included is much larger than
any other energy scale in the problem, then the error due to truncation is
small.

\subsection{Coupling to leads}

The transport  properties of the shuttle system coupled to the leads are
obtained by integrating an equation of motion for the density matrix
appropriate to an open quantum system. Apart from the internal dynamics of the
shuttle, there are two effects we need to take account of. Couplings between
the electronic states in the outer dots and the leads must be incorporated,
and the coupling between the oscillator and its environment must also be
included.

The external couplings of the outer dots  and oscillator are accounted for in
the dynamics via additional terms in the equation of motion for the reduced
density matrix of the shuttle, an approach which is well--known in the field of
quantum optics and has recently been applied to problems of electron transport
in nanostructures.\cite{dd1,gurvitz}
The electrons in the leads are assumed to
be completely incoherent which means that all the off-diagonal elements
of the density matrix between the state $|0\rangle$ and the other electronic
states can be set to zero (notice, however, that the diagonal element in the
density matrix for $|0\rangle$ has both diagonal and off-diagonal
matrix elements in the space of the vibrational mode states).

The appropriate equation of motion for the density matrix of the system
generalised to include the leads and the environment of the oscillator
therefore takes the general form
\begin{equation}
\dot{\rho}=-\frac{i}{\hbar}[H,\rho]+\Xi\rho+\dot{\rho}_d
\end{equation}
where the `decay matrix', $\Xi$, incorporates transitions between the leads and the outer
dots and $\dot{\rho}_d$ accounts for the decohering and dissipative effects of
the
oscillator's environment. The form of the non-unitary terms can be derived by
adapting standard techniques used in the field of quantum optics and we will
consider them both in turn.\cite{qo}

The form of decay matrix coupling the shuttle to the leads can be obtained
using the method of  Gurvitz {\emph{et al.}},\cite{gurvitz} as described in
appendix B. The components of the matrix, $[\Xi\rho]^{ij}_{ab}$,
are labelled by the vibrational state (superscripts) and the electronic
states (subscripts). The matrix is Hermitian and is specified by the
components\cite{mat}
\begin{eqnarray}
\left[\Xi\rho\right]^{ij}_{ll}&=&\rho_{00}^{ij}\Gamma \label{one}\\
\left[\Xi\rho\right]^{ij}_{rr}&=&-\rho^{ij}_{rr}\Gamma \\
\left[\Xi\rho\right]_{00}^{ij}&=&-\Gamma(\rho_{00}^{ij}-\rho^{ij}_{rr}) \\
\left[\Xi\rho\right]^{ij}_{rc}&=&-\frac{\Gamma}{2}\rho^{ij}_{rc} \\
\left[\Xi\rho\right]^{ij}_{lr}&=&-\frac{\Gamma}{2}\rho^{ij}_{lr} \\
\left[\Xi\rho\right]^{ij}_{cc}&=&\left[\Xi\rho\right]^{ij}_{lc}=0, \label{two}
\end{eqnarray}
where $\Gamma$ is the tunneling rate between the dots and the
leads, and all terms $\left[\Xi\rho\right]^{ij}_{0a}$ where $a\neq
0$ are zero as the associated elements of the density
 matrix ($\rho^{ij}_{0a}$ with $a\neq 0$) are zero by definition.

Including the environment of the vibrational mode is essential to the
description as it is
the dissipation arising from this coupling which gives rise to
a steady-state in which the current is constant, as we discuss below.
We employ a minimal model of the environment, assuming it to
be composed of a bath of oscillators at a fixed temperature $T$, to which the
vibrational mode is coupled only weakly. Under these assumptions the dissipative
component in the equation of
motion for the density matrix, is given by\cite{qo}
\begin{eqnarray}
\dot{\rho}_d&=&-\frac{\gamma}{2}{\bar {n}}(\hat{d}\hat{d}^{\dagger}\rho
-2\hat{d}^{\dagger}\rho \hat{d}
+\rho \hat{d}\hat{d}^{\dagger})\\ \nonumber
&&-\frac{\gamma}{2}({\bar {n}}+1)
(\hat{d}^{\dagger}\hat{d}\rho-
2\hat{d}\rho \hat{d}^{\dagger}+\rho \hat{d}^{\dagger}\hat{d}),
\end{eqnarray}
where $\gamma$ is the classical damping rate of the oscillator, and $\bar{n}$,
is the usual thermal occupation number of the oscillator at temperature $T$,
\begin{equation}
{\bar{n}}=\frac{1}{{\rm e}^{\hbar\omega/k_{\rm B}T}-1}.
\end{equation}
The classical damping rate is just the rate at which the vibrational mode loses
energy due to frictional forces, its value can be obtained empirically
from the quality factor of the oscillator,\cite{qo,jn2} $\gamma=\omega/Q$.

The steady-state current through the system (in units of electrons per unit
time) is given by
\begin{equation}
\frac{I}{e}=\Gamma\rho_{rr}^{(s)},
\end{equation}
where $\rho_{rr}^{(s)}$ is the occupation probability of the righthand
dot when a steady state has been achieved. In practice the current is
determined after evolving the equation of motion for the density matrix
until further changes with time become negligible.

The discrete nature of the states in the dots has an important effect on the
ways in which energy can be transferred in the shuttle. An electron travelling
through the shuttle must enter at an energy determined by the level in
the lefthand dot and leave at the energy determined by the lower level in the
righthand dot. Hence each electron travelling through the device dissipates
an amount of energy proportional to the bias voltage. For a system without an
oscillator,\cite{three} the energy is dissipated in the lead with the lower chemical potential.
However, for the shuttle we consider here, some of the energy ends up in the
oscillator---rather like the electron `pump' considered by Stafford and Wingreen
in reverse.\cite{dd2} Despite this apparent pumping mechanism for the oscillator,
the degree to which the oscillator can be excited is strongly limited
by damping and a steady-state is always achieved when this effect is
included.\cite{note3}

\section{Eigenvalue dynamics of the isolated shuttle}

For a system in which the coupling to external leads and the
environment\cite{three} is much weaker than the coupling between the
dots (i.e.\ when $\Gamma, \gamma\ll V{\rm e}^{-\alpha x_0}/\hbar$), the
current characteristics are expected to be strongly
influenced by the eigenstates of the isolated system. Under these conditions,
we can think of the eigenstates of the system, which in general are not
localised on any one dot, as forming independent channels for
conduction.  The current is carried most effectively by those eigenstates
in which there is a finite probability of finding an
electron in both the left and righthand dots, and the characteristics
of the eigenstates  can be determined from the behavior of the
corresponding eigenvalues. Therefore we examine dynamics of the eigenvalues
as a function of the bias voltage in the uncoupled shuttle system before going
on to consider the corresponding  behavior of the current in the coupled
system.

The eigenvalue spectrum of the shuttle at zero bias voltage is
controlled by the relative values of the energy scales of the oscillator,
$\hbar\omega$,  and the tunneling matrix element, $V$; and by the
length-scales $x_0$ and  $1/\alpha$. The eigenvalue dynamics in the limit
where the dots are uncoupled (i.e.\ $V=0$) is illustrated in Fig.\ 2a [note
that all the energies in the  figures are measured in units of $\hbar\omega$,
and the distances are measured in units of $\Delta x_{zp}$]. The energy
levels at zero-bias  are three-fold degenerate  states separated by
$\simeq\hbar\omega$. For small  applied bias, the splittings between energy
levels are linear. The spacing between the sets of three levels at zero-bias
are not exactly $\hbar\omega$, but increase slightly with energy. This is
because the  oscillator is in fact confined by hard wall potentials which it
begins to feel at higher  energies, leading to an increase in the
eigenenergies.

When the electronic states are decoupled the individual dot states are
readily identified: the left and righthand dot energy levels increase and
decrease  respectively with increasing voltage bias, whilst the energy of the
central-dot electronic states drop off quadratically
(this state is equivalent to an oscillator in a linear potential).
The eigenvalues for the left and righthand energy levels cross for
$\varepsilon_{b}\simeq n\hbar\omega$ with $n$ an integer.
For $\varepsilon_{b}\simeq 2\hbar\omega$, the crossing of
the left and righthand-dot energy levels almost coincides with the
central-dot energy level. The level crossings of the lower
eigenvalues occur at almost precisely integer values of $\hbar\omega$, but at
higher energies the level crossings drift to larger bias values because of the
increased spacing between levels at zero-bias.

For finite values of the tunneling matrix element, $V$, the eigenvalue
dynamics as a function of $\varepsilon_{b}$ becomes more complex, as
shown in Fig.\ 2. The energy levels at  zero-bias split so that the degeneracy
of the levels is lifted. At finite bias the levels no longer cross, but instead repel each
other. The strength of the level repulsion depends strongly on $V$,
but  the oscillator state also plays a role as
larger number states for the oscillator lead to stronger couplings
between levels. However, away from  the anti-crossings, the levels
vary with $\varepsilon_b$
in almost the same way as for the decoupled case, implying that the eigenvalues
in these regions can still be associated with states localised on the
individual dots.

The level repulsions which occur at $\varepsilon_b\simeq n\hbar\omega$ for
finite $V$ have a different character for odd and even values of $n$. For odd
values of $n$, levels
associated with  the left and righthand dots anti-cross, leading to mixing
of the levels. For even values of $n$, levels
associated with the central dot mix first with levels associated with the
lefthand dot and then with levels associated with the righthand dot within a
very narrow range of $\varepsilon_b$. The simultaneous curvature of the three
levels implies that, over a narrow range
of $\varepsilon_b$, the mixing involves all of them. The lefthand dot level
mixes with the central dot level first partly because
the central dot level curves downwards, and partly because the lefthand dot
level is associated with a less excited state of the oscillator which lies
closer to the central dot level at zero-bias than the righthand dot level.

In the vicinity of the level anti-crossings states associated with either the
lefthand and righthand dots, or with all three dots, become strongly mixed.
The mixing of these levels implies that the electron will be highly delocalised
near the anti-crossings, leading to a strong enhancement of the current
through the shuttle. The variation in
the location and strength of anti-crossings of levels (and hence mixings of
states associated with different dots) with the eigenvalue energy implies that
the actual current characteristics will be sensitive to the
energy of the shuttle in the steady state.

At zero bias the system will be at resonance as the dot energy levels will be aligned,
and we expect a large current to flow.
However, the large current at zero bias can also be understood in terms of the
current channels formed by the eigenstates.
At very small bias energies the eigenvalues for finite
$V$  have a quadratic dispersion, and the eigenstates are composed of
almost symmetrical mixtures of the lefthand, central and righthand dot states
and hence are highly delocalised.\cite{sym}

In studying the eigenspectra and the steady-state current we concentrate on
a range of bias values of a few $\hbar\omega$, for a combination of reasons.
As the bias voltage increases, the eigenspectra become more complex. The
spread in the locations of the anti-crossings with eigenenergy increases
with $\varepsilon_b$ and the energies of the levels which anti-cross increase.
The complexity of the eigenspectra at large values of $\varepsilon_b$  means
that it will become increasingly difficult to understand the current
characteristics of the system by reference to the eigenspectrum.
Furthermore,  our simple modelling of the electronic part of the system,
including only a single energy level for each dot, will become increasingly
unreliable as the bias is increased and higher levels in the dots become
accessible.

\section{Current characteristics}

In order to obtain the current characteristics of the shuttle system we
integrate the equation of motion for the density matrix from a given initial
state, until a steady state is obtained. The steady state is achieved when the
energy gained by the oscillator from the electrons is matched by losses due
to damping by the surrounding medium.

The steady-state current for a chain of quantum dots at fixed
positions has been studied
extensively.\cite{three,dd1a,extra,angus1,angus2} In the Coulomb
blockade regime, it is found that if the energy level in each of
the dots is shifted by an amount proportional to its position
(forming a `Stark ladder'), the decay in current with the shift in
energy levels is rapid.\cite{three} For the shuttle, we find that
the presence of the oscillator leads to  significant changes in
the current characteristics.

We begin by analysing the simplest case, where the dots are
strongly coupled to each other (and weakly coupled to the leads)
and  the temperature is set to zero. In this regime the current
characteristics can be understood in terms of the energy
eigenspectrum of the system and the effect of the oscillator's
environment which causes scattering between the eigenstates.
However, we also consider the behavior when the tunneling length
of the electrons is reduced, so that the coupling between the dots
becomes weak,  as it is in this regime that current can flow by
hopping onto then off the central dot sequentially, in a manner
very reminiscent of  the semi-classical
shuttles.\cite{Gk,blick,blick2,weiss} Furthermore, in practice,
temperature will play an important role and so we extend our
analysis to small but finite temperatures. Working at finite
temperatures complicates the numerical calculation as it naturally
increases the importance of higher energy oscillator states. This
means that care needs to be taken in truncating the matrix of
oscillator states (as described in appendix A) to ensure that the
results obtained do not depend strongly on the size of the
truncated Hamiltonian matrix. The errors due to truncation only
become significant for temperatures such that $\hbar\omega\ll
k_{\rm B}T$. The matrix method can readily be adapted to
investigate the effect of temperature when $\hbar\omega\sim k_{\rm
B}T$.

The initial state of the system
describes the initial wavefunctions of the electronic  and oscillator parts
of the system. The initial form of electronic part of the wavefunction is
unimportant:
it simply describes the initial probabilities of finding an electron on each
of the dots, which does not in practice affect the form of the steady state which is
later achieved. However, the initial state of the oscillator is of some
importance as it depends on the background temperature, but  at
zero temperature, the oscillator is initially in its ground state.

The steady-state current through the dots as a function of $\varepsilon_b$
with and without the oscillator is illustrated in Fig.\ \ref{fig:three} for the
case of strong coupling between the dots.
The tunneling rate to the leads is $\Gamma=0.05$ and the damping rate for the
quantum oscillator is $\gamma=0.025$ [all rates are measured in units of
$\omega$].

When the bias is zero, all three dot levels are aligned leading to
resonant transmission of electrons. Without
the oscillator, the current decays rapidly with increasing bias,
$\varepsilon_b$, as expected. When the oscillator is present,
the current initially decays with increasing bias, though more slowly than
before,  but at larger values of $\varepsilon_b$ prominent current
resonances dominate the behavior.

For strong coupling between the dots (compared to the couplings to
 the leads), the current peaks are best
described in terms of the conduction channels formed by the eigenstates of the
shuttle system. Comparison of the current peaks in Fig.\ 3 with the corresponding
eigenspectrum in Fig.\ \ref{fig:two} shows that the resonances
coincide with avoided-level crossings where the eigenstates are formed from
mixtures of states localised on the individual dots.  Further resonances
should occur at higher values of $\varepsilon_b$, but the model will eventually break down
 when the shift in levels exceeds the energy level spacing in the dots
 themselves. The peak in the current at zero bias arises because the
eigenstates for $\varepsilon_b=0$ each consist of (symmetrical) mixtures of the
states associated with each of the dots. Of course the peak at zero bias
can also be seen as due to resonant tunneling through the shuttle.

The second current peak, which occurs at $\varepsilon_b\simeq 0.86$ (in units
of $\hbar\omega$), is
due to  the mixing of levels associated with the left and righthand dots. The
resulting eigenstates are symmetrical as they contain almost equal weights for the left and righthand
dots (i.e.\ in these states the probability of finding the electron on the
left and righthand dots are are almost equal).
The third current peak occurs at $\varepsilon_b\simeq 1.95$, and corresponds
to a mixing of the central dot level, more or less simultaneously, with both
left and righthand dot levels. In this case the resulting eigenstates are not
entirely symmetrical: there are larger weights for the lefthand dot in some
states and larger weights for the righthand dots in others.  The fourth current
peak, occuring at    $\varepsilon_b\simeq 2.92$, is similar to the second
peak, in that it too is due to the mixing of levels associated with the left
and righthand dots. The magnitude of the fourth current peak is rather lower
than that of the second peak, partly because the left and righthand dot levels
which are  mixed (at  $\varepsilon_b\simeq 2.92$),  correspond to oscillator
states differing by three quanta (whereas states differing by only one
oscillator quantum are mixed at $\varepsilon_b\simeq 0.86$), and  partly
because anti-crossings progressively disappear from the lower part of the
energy spectrum as the bias is increased.

\subsection{Effects of oscillator damping}
The dynamics of the system is strongly influenced by the
magnitude of the oscillator damping, $\gamma$. Without
any damping, the oscillator would gain energy continuously from the
electrons and no steady state would
be achieved (without including higher electronic levels).
The effect of $\gamma$ in producing a steady state can be seen
by examining the average energy of the shuttle
as a function of time.

The average energy is readily computed from the
density matrix and Hamiltonian of the system (suitably generalised to account
for the energy of the oscillator when none of the dots are occupied).
Fig.\
\ref{fig:threex} shows the evolution of the average energy, $\langle E\rangle
={\rm Tr}[H\rho]$, as a function of
time for a range of different damping constants. The average energy at $t=0$
is given by the initial state of the system with the oscillator in its ground
state and an electron present on the lefthand dot (the bias energy,
$\varepsilon_b=0.86$, corresponds to a peak in the current).
For each value of $\gamma$, the average energy varies with time before
eventually settling down to a constant value, indicating that the system has
reached a steady-state.

The time taken to reach the steady-state, and the average energy in that state
vary considerably. For $\gamma\sim 0.025$ the oscillator appears to be
`critically damped' in the sense that the energy in the steady-state  is
roughly the initial energy of the system. In contrast, for $\gamma > 0.025$
the system achieves a steady state by losing energy to the environment,
implying over-damping of the oscillator.  As soon as the electron leaves the
lefthand dot and moves to states on the other dots at
lower energies, some of the initial energy is dissipated into the environment
via the oscillator. When $\gamma<0.025$, the oscillator is under-damped and the
system gains energy from the electron reservoirs as successive electrons pass
through the  system until a steady state is eventually achieved. For the
underdamped case, the time taken to achieve the steady-state is longer than for
either  critical or over-damping and it eventually diverges when $\gamma$ is set
to zero.

The value of the oscillator damping also has a strong effect on the
magnitude of the steady-state current
through the shuttle, as shown in Fig.\ \ref{fig:fiver}. The way in which
 $\gamma$ affects the current depends sensitively on the value
of $\varepsilon_b$. The first and second current peaks are
almost unaffected by the variation in the oscillator damping constant, $\gamma$,
whilst damping of the oscillator increases the current at the third peak and
decreases it at the fourth peak.
The reason for this differing behavior is readily understood in terms of a
picture of independent conduction channels arising from the eigenstates of
the shuttle. If the damping is weak, it can be  thought of as
causing scattering between channels with slightly differing energies.
Therefore, damping can increase the current wherever scattering can transfer
electrons from channels with a larger  weights on the lefthand dot (where the
electrons enter the system) to ones with large weights on the righthand dot
(from which the electron leaves the system). At the first and second current
peaks
the states which carry the current are largely symmetrical, in that they
contain almost equal weights for the left and righthand dots, so the
damping cannot increase the current. In contrast, at the third peak the
current can increase as there is considerable asymmetry between nearby
eigenstates.  At the fourth current peak, increasing the damping reduces the
current because at this bias energy  the current is very
sensitive to the reduction in the average energy caused  by the higher
damping.

\subsection{Current in weak coupling regime}

At larger values of $\alpha$ (i.e.\ for smaller tunneling lengths $\lambda
=1/\alpha$) the coupling between the dots grows weaker and the
picture of the current as arising from the passage of electrons through weakly
coupled eigenstates acting as conduction channels begins to break down. Instead
we must think of the electrons as tunneling between states localised on the
individual dots.

When the bias energy is $\sim \hbar\omega$ or $3\hbar\omega$  the electrons
must tunnel coherently from the lefthand dot to the righthand dot via a virtual
(non-energy conserving) transition to the  central dot.
However, when the bias energy is $\sim 2\hbar \omega$ electrons can tunnel
through the system incoherently via two separate tunneling events in each of
which a quanta of energy is dumped in the oscillator. Of course when
$\varepsilon_b=0$ there is no need to dump energy
into the oscillator and all tunneling processes through the central dot
conserve energy.

The effect of increasing $\alpha$ is shown clearly in  Fig.\
\ref{fig:alpha}: the second and fourth peaks are strongly
suppressed, whilst the  third peak is only slightly reduced. As
the tunneling length becomes smaller, coherent tunneling directly
between the left and righthand dots becomes more difficult.
However, electrons can still flow through the device by tunneling
onto the central dot and then off again later in a sequence of two
separate tunneling events, i.e.\ current flows by shuttling of
electrons through the central dot.

As for strongly coupled dots, an increase in damping of the oscillator
enhances the current when $\varepsilon_b\simeq 2\hbar\omega$. When the
coupling between the dots is weak the damping assists current flow by removing
energy from the oscillator and thereby helps prevent electrons from
tunneling backwards (i.e\ from the righthand dot to the central dot or from the
central dot to the righthand dot).

\subsection{Finite temperatures}

The effects of finite temperature on the current characteristics of the shuttle
can be taken into account by generalising the initial state of the oscillator
to a thermal mixture and modifying the damping terms (described in section II)
to take account of the environment's temperature. As long as the temperature
does not approach the charging energies associated with the Coulomb blockade
on the dots, no adjustment to the electronic part of the model is required.

Fig.\ \ref{fig:four} compares the current through the shuttle for a particular
choice of parameters at $k_{\rm B}T=0$ and $k_{\rm B}T=3$ (in units of $\hbar \omega$). The finite temperature has
two noticeable effects. Firstly, the current characteristics are smeared out:
the peaks are broader and lower, while the troughs are shallower. Secondly, the
maxima of the peaks are shifted to slightly higher bias energies.

The effect of finite temperature on the current is readily interpreted
with reference to the eigenspectra in Fig.\ \ref{fig:two}. The finite temperature
increases the average energy of the system, whilst also broadening the
distribution of probabilities of finding the system in a given state. The
broadening of the probability distribution implies that a whole range of
states contribute to the current, each of which have anti-crossings occuring
at different values of $\varepsilon_b$, leading to a broadening of the current
peaks. The overall increase in the average energy leads to a general
increase in the current because of the increased strength in the anti-crossings
at higher energies, especially at larger values of  $\varepsilon_b$ where
anti-crossings no longer occur in the lower parts of the eigenspectrum.

\section{Conclusions and discussion}

We have investigated the effect of a quantised vibrational degree of freedom
on transport through a chain of three quantum dots where the tunneling between
the dots depends exponentially on the displacement of the oscillator. When
the dots are strongly coupled to each other and weakly coupled to the leads,  the
current through the chain can be predicted by looking at where anti-crossings
in the eigenspectrum give rise to electronic states which are delocalised
amongst the dots.

The damping of the oscillator due to coupling to its environment has
important effects on the current flowing through the system.
The current through the shuttle only reaches a steady value when the energy
pumped into it from the electrons is balanced by the dissipation
due to the oscillator's environment. For weak damping, the environment can
be thought of as causing scattering between different current channels. The
scattering enhances the current whenever it can transfer electrons from states
strongly concentrated on the lefthand dot to ones strongly concentrated on the
righthand dot. However, when the eigenstates contain equal probabilities of
finding the electron on the lefthand and righthand  dots, scattering has no
effect on the current.

When the tunneling length  is reduced the effective coupling
between the dots is also reduced and the picture in terms of
independent conduction channels can no longer be applied. Instead
we can think of the electrons as passing through the shuttle via
transitions between states localised on each of the dots. When the
energy difference is an odd number of vibrational quanta, current
flow is via a virtual transition. In contrast, when the energy
difference is an even number of vibrational quanta, electrons can
tunnel sequentially onto  and then off the central dot. The
virtual transitions are strongly suppressed by reductions in the
tunneling length of the electron. However, the  sequential
transitions---electrons hopping on and then off the central
dot---are much less affected by changes in the tunneling length.

The transport of electrons through the central dot via sequential
tunneling, is analogous to the semi-classical models of
electron-shuttling.\cite{weiss,Gk} As in the case of the
experiments of Blick {\it et al.},\cite{blick,blick2} the
vibrational mode is driven and current flows by electrons
tunneling onto the central dot and then tunneling off in the
opposite direction at a later time. However,  for the system we
consider the driving energy comes from the electrons themselves as
the difference in dot energy levels forces them to lose energy to
the oscillator.

The shuttling mechanism dominates over higher order processes when
the tunneling length, $\lambda$, is of order the zero-point
position uncertainty of the oscillator, $\Delta x_{zp}$, and can
give rise to a larger current than at the electronic resonance
(i.e.\ when $\varepsilon_b=0$). However, the sharp features in the
current characteristics begin to be smeared out when $k_{\rm
B}T>\hbar\omega$.

In order to have $k_{\rm B}Tsim\hbar\omega$ in the range 0.1--1K
the frequency of the oscillator must be  $\sim1$--10GHz. This high
frequency range is accessible in a wide variety of systems: the
elastic links in nanopartice arrays have frequencies\cite{nish}
$\sim10$GHz, micromechanical resonators can be fabricated with
frequencies approaching 1GHz,\cite{ghz} and indeed the vibrational
modes in many molecular systems have much higher
frequencies.\cite{nat} The tunneling length depends on the work
function of the surfaces involved, but typically lies in the
range\cite{Gk} $0.05$--$3$
 \AA. Unfortunately, raising the frequency of the oscillator
 reduces
the  zero-point position uncertainty, $\Delta
x_{zp}=\sqrt{\hbar/2m\omega}$, but the effective masses of even
artifically fabricated oscillators can be extremely small (for
example nanoparticles can have masses\cite{nish} $\sim
10^{-23}$kg). Therefore, it should be possible to fabricate or
assemble systems in which both $\lambda \sim\Delta x_{zp}$ and
$k_{\rm B}T\sim\hbar\omega$, and thereby observe electron
shuttling in the quantum regime.

One of us (A.M.) would like to thank The Cavendish
Laboratory, at the University of Cambridge, for their hospitality. This work
was funded by the EPSRC under grant GR/M42909/01.

\appendix
\section{Hamiltonian in matrix form}
The Hamiltonian of the isolated system of three dots and an oscillator has three
electronic states and a number of vibrational states which is in principle
infinite, but which in practice we limit to a number $N\sim 25$ so that
the largest energy state of the oscillator ($E_{max}=N\hbar\omega$) is much
greater than any other energy scale in the problem.

The full Hamiltonian for the shuttle consists of nine $N\times N$ sub-matricies, corresponding
to the three electronic states and the tunneling elements between them
\begin{equation}
H=\left( \begin{array}{ccc}
H_{LL}&H_{LC} &H_{LR} \\
H_{CL} & H_{CC}& H_{CR} \\
H_{RL}& H_{RC} & H_{RR}\end{array}\right).
\end{equation}
The sub-matrices have rows and columns labelled in a suitable basis of the
the oscillator states. In the number representation of the oscillator states
the diagonal sub-matrices take the form
\begin{eqnarray}
\langle n| H_{LL}|m\rangle&=&(\varepsilon_l +m\hbar\omega)\delta_{nm} \\
\langle n| H_{CC}|m\rangle&=&\frac{\varepsilon_d}{\Delta x_{zp}}\langle n|(\hat{x}+x_0)
|m\rangle \nonumber \\
&&+(\varepsilon_l+m\hbar\omega)\delta_{nm} \\
\langle n| H_{RR}|m\rangle&=&(\varepsilon_r +m\hbar\omega)\delta_{nm}, \\
\end{eqnarray}
where $\varepsilon_d=\Delta x_{zp}(\varepsilon_r-\varepsilon_l)/2x_0$,
with $\Delta x_{zp}$ the zero-point position uncertainty of the oscillator.
Of the other sub-matrices,
only those coupling adjacent dots (i.e.\ $H_{LC}$ and $H_{CR}$, and their
Hermitian conjugates) are non-zero,
\begin{eqnarray}
\langle n| H_{LC}|m\rangle&=& -V\langle n|\mathrm{e}^{-\alpha
[x_0+\hat{x}]}|m\rangle \\
\langle n| H_{CR}|m\rangle&=& -V\langle n|\mathrm{e}^{-\alpha
[x_0-\hat{x}]}|m\rangle.
\end{eqnarray}

The most efficient representation of the oscillator states for computational
purposes is the position basis. The oscillator displacement is restricted to
lie between the values set by the positions of the outer dots, a restriction
which is readily enforced in the position representation. The set of
position states is obtained by diagonalising the position operator in the
number basis (a matrix of size $N\times N$), then those states amongst
the resulting $N$ position states with eigenvalues outside
the range $(-x_0,x_0)$ are projected out, leading to a somewhat smaller
sub-matrix. This simple-minded procedure is appropriate at relatively
low energies when the weight of the wavefunction in the region outside
the cut-offs is relatively small.

\section{Density matrix equation of motion}
In this appendix we detail how the dissipative terms in the
density matrix equation of motion are obtained when the voltage
difference between the leads is less than the spacing of energy
levels within each of the dots, so that transport is restricted to
the Coulomb blockade regime. We begin by deriving the rates which
describe the transitions to and from the leads which are attached
to the system, before going on to calculate the equation of motion
for the  density matrix following the prescription of Gurvitz
{\emph{et al.}}\cite{gurvitz}

Following the usual theory of single electron tunneling\cite{bean}
we can write the net tunneling rate from a lead with chemical potential
$\mu$ through a barrier to an isolated state with effective energy $\varepsilon$ as
\begin{equation}
\Gamma_{net}=V_0\left[P(0)f(\Delta)-P(1)(1-f(\Delta))\right]
\end{equation}
where $\Delta=\varepsilon-\mu$, $V_0$ is the tunneling amplitude
which is assumed to be independent of the energy, $f$ is the Fermi
distribution function
 and $P(0)$($P(1)$) is the probability that the
isolated state is unoccupied(occupied).

For the shuttle, we need to consider transitions between the
lefthand lead and lefthand dot, and between the righthand lead and
righthand dot. In addition, the strong Coulomb blockade means that
transitions onto any of the dots can only occur if all of the dots
are unoccupied. Therefore, for transitions between the lefthand
lead and left-hand dot, $P(0)=\rho_0$ where we use the notation
$\rho_{k}=\sum_i \rho^{ii}_{kk}$, for the probability that the
electronic state $k$ is occupied, with the superscripts and
subscripts of the density matrix labelling the oscillator and
electronic states respectively. Thus the net tunneling rate from
the left-hand lead to the left-hand dot takes the form
\begin{equation}
\Gamma^L_{net}=\Gamma_L^+\rho_0-\Gamma_L^-\rho_l \label{x}
\end{equation}
where $\Gamma_L^+=V_0 f(\Delta)$, $\Gamma_L^-=V_0(1-f(\Delta))$.
Similarly, we find
\begin{equation}
\Gamma^R_{net}=\Gamma_R^+\rho_r-\Gamma_R^-\rho_0.
\end{equation}
These net rates can be separated into individual terms which enter into the
equations of motion of the elements of the density matrix. We assume that
$\Delta=\varepsilon_l-\mu_l=\mu_r-\varepsilon_r$, where $\mu_{l(r)}$ is the Fermi
level in the left(right) lead. The tunneling rates for the righthand lead
take the form $\Gamma_R^-=V_0 f(-\Delta)$ and $\Gamma_R^+=V_0(1-f(-\Delta))$.

In order to simplify the analysis, we will work in the large bias limit
 where $|\Delta|\gg k_{\mathrm{B}}T$. In this limit, the backward tunneling
 rates
 $\Gamma_{L(R)}^-$ go to zero and the forward rates take the simplified form
 $\Gamma_{L(R)}^+=V_0$. We shall restrict our analysis to this regime for
 simplicity, but the generalisation to the master equation for the case where
 $|\Delta|\sim k_{\mathrm{B}}T$ is relatively straightforward.

Gurvitz {\it et al.},\cite{gurvitz} have developed a way of
extending the density matrix description of electronic
nanostructures to  include couplings between localised, coherent,
electronic states and leads containing a very large number of
states where the electrons can be regarded as incoherent.
According to this approach, the equation of motion for the density
matrix including couplings to the leads (but not the environment
of the oscillator) takes the general form\cite{simplify} (setting
$\hbar=1$)
\begin{eqnarray}
\dot{\rho}_{aa}^{ij}&=&-i[H,\rho]_{aa}^{ij}-\rho_{aa}^{ij}\sum_{d\neq a}
\Gamma_{a\rightarrow d}+\rho_{cc}^{ij}\sum_{c\neq a}\Gamma_{c\rightarrow a}
\label{gu2} \\
\dot{\rho}_{ab}^{ij}&=&-i[H,\rho]_{ab}^{ij}-\frac{\rho_{ab}^{ij}}{2}\left(
\sum_{d\neq a}\Gamma_{a\rightarrow d}+\sum_{d\neq b}\Gamma_{b\rightarrow d}
\right)   \label{gu}
\end{eqnarray}
where the superscripts on the density matrix label the vibrational
state and the subscripts label the electronic parts. The terms
$\Gamma_{n\rightarrow m}$ give the transition rates from states
$|n\rangle$ to $|m\rangle$.

Following the prescription given in Eqn.s (\ref{gu2}) and
(\ref{gu}), and taking into account only tunneling onto the
left-hand dot from the left-hand lead and
 tunneling off the righthand dot into the righthand lead, we obtain
\begin{eqnarray}
\dot{\rho}_{ll}^{ij}&=&-i[H,\rho]_{ll}^{ij}+\rho_0^{ij}\Gamma_L\\
\dot{\rho}_{cc}^{ij}&=&-i[H,\rho]_{cc}^{ij} \\
\dot{\rho}_{rr}^{ij}&=&-i[H,\rho]_{rr}^{ij}-\Gamma_R\rho_{rr}^{ij} \\
\dot{\rho}_0^{ij}&=&-i[H,\rho]_{00}^{ij}-\Gamma_L\rho_0^{ij}+\Gamma_R\rho_{rr}^{ij} \\
\dot{\rho}_{lc}^{ij}&=&-i[H,\rho]_{lc}^{ij} \\
\dot{\rho}_{lr}^{ij}&=&-i[H,\rho]_{lr}^{ij}-\frac{1}{2}\Gamma_R\rho_{lr}^{ij} \\
\dot{\rho}_{cr}^{ij}&=&-i[H,\rho]_{cr}^{ij}-\frac{1}{2}\Gamma_R\rho_{cr}^{ij},
\end{eqnarray}
where we have omitted the superscript `$+$' from the tunneling rates and
the terms off-diagonal in the electronic state $|0\rangle$ are zero, by
definition, and so are omitted. Assuming the
dot-lead junctions are identical, we can simplify the equations further by writing
$\Gamma=\Gamma_R=\Gamma_L$, and the terms coupling the dots to the leads are
given by the matrix, $\Xi\rho$, given in Eqn.s \ref{one}--\ref{two}.

\newpage

\begin{figure}
\center{
\epsfig{file=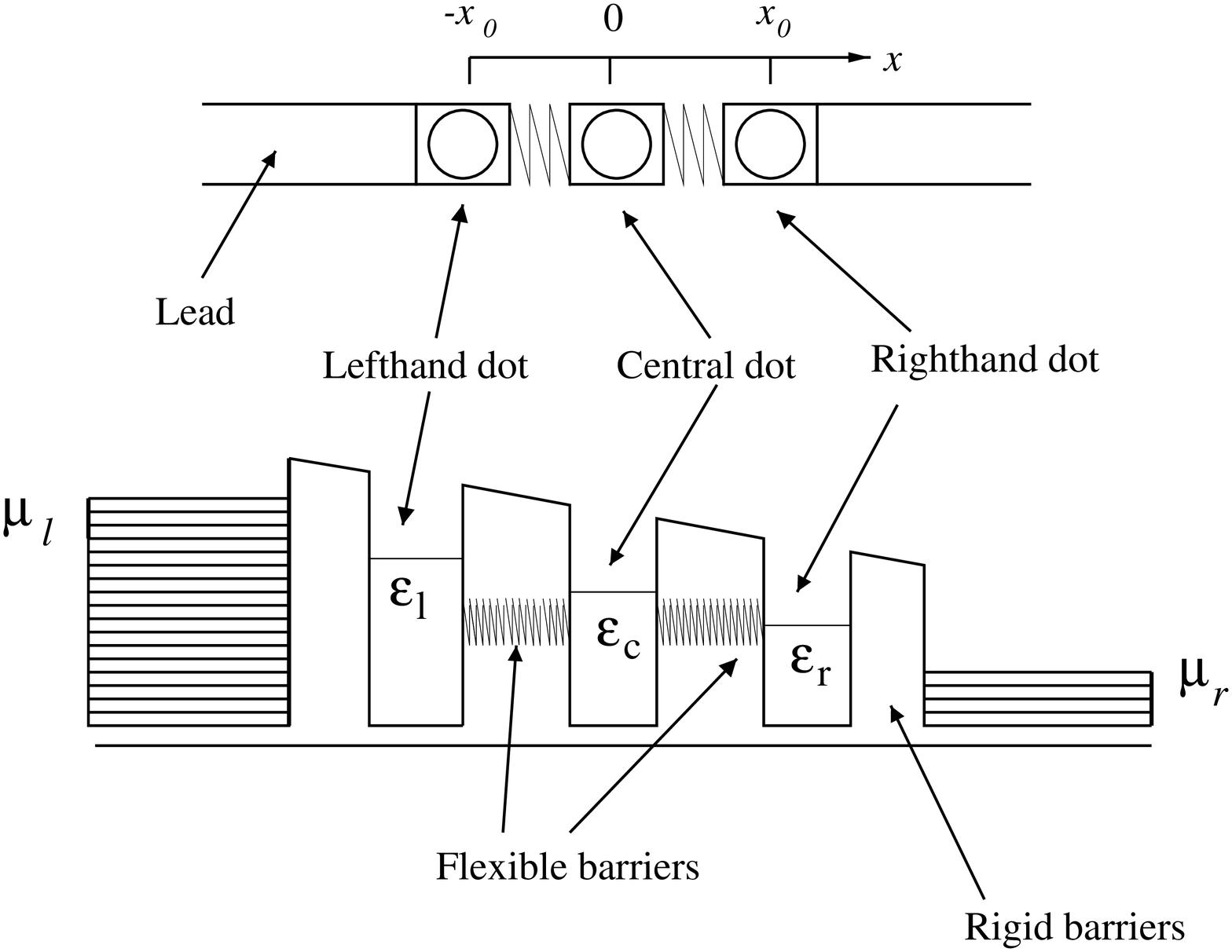, width=12.0cm}}
\caption{\small{Schematic diagrams of the shuttle system. The
geometry and band structure of the system are shown in the upper
and lower panels respectively.}}
\label{fig:one}
\end{figure}

\begin{figure}
\center{
\epsfig{file=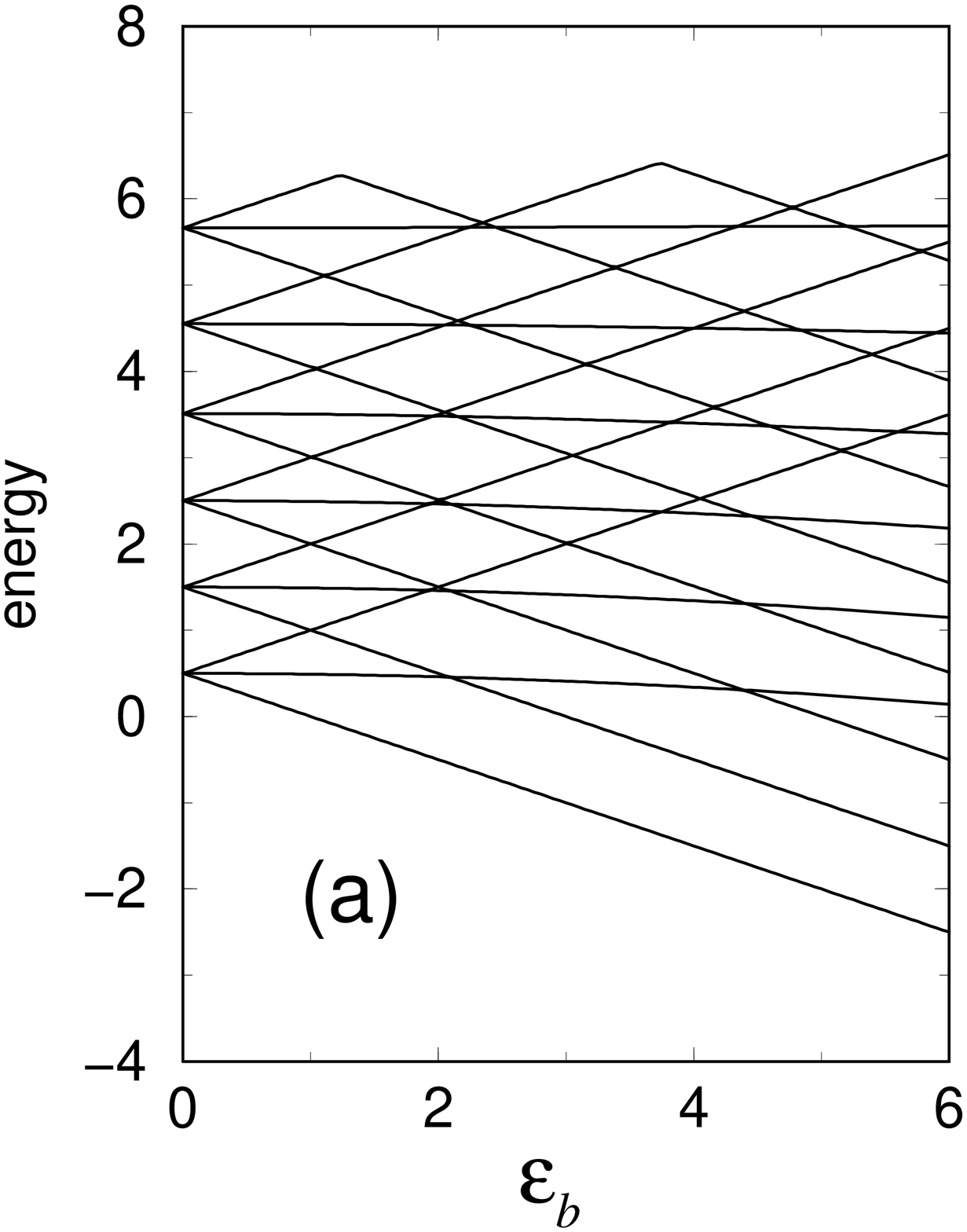, width=6.0cm}
\epsfig{file=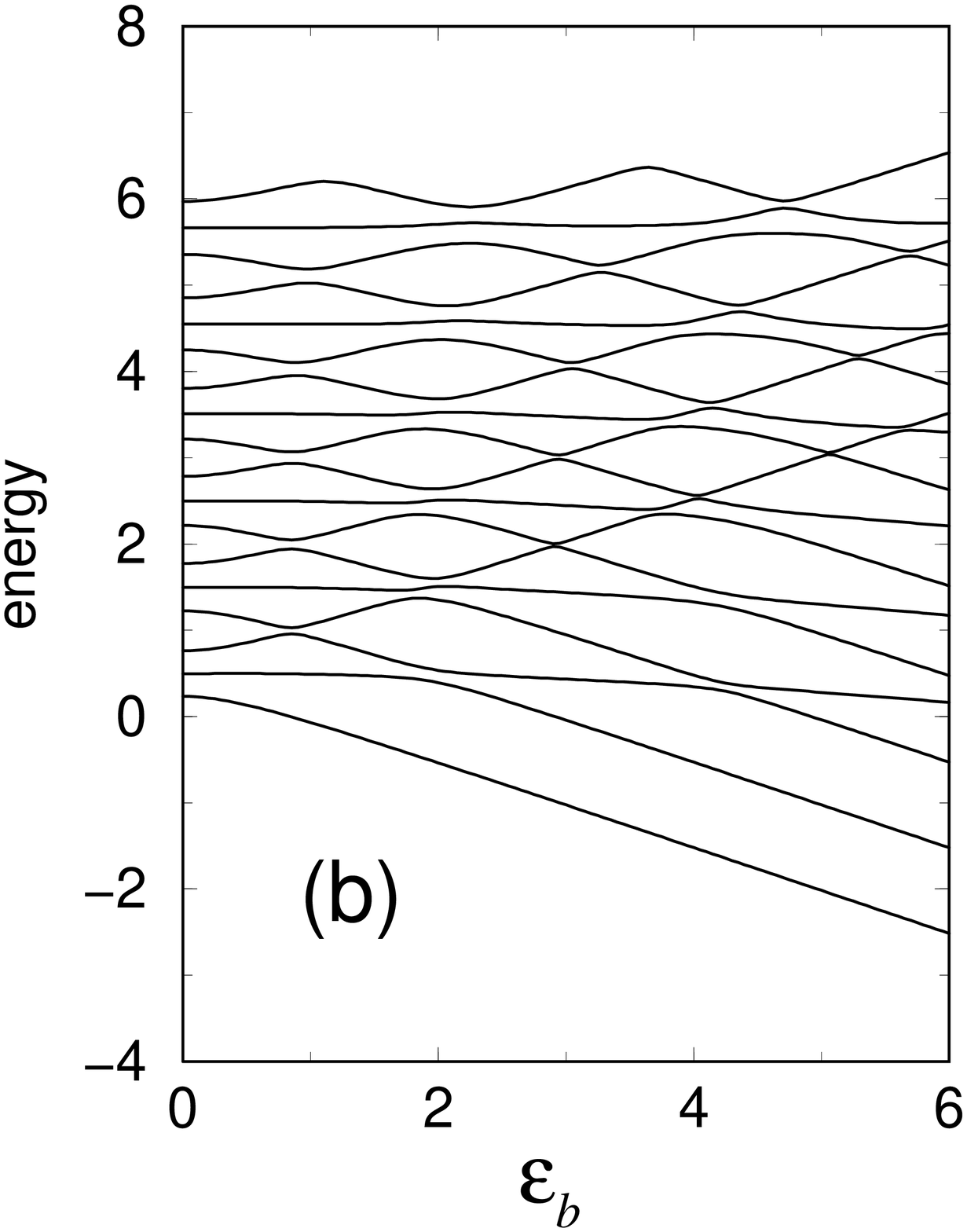, width=6.0cm}
\epsfig{file=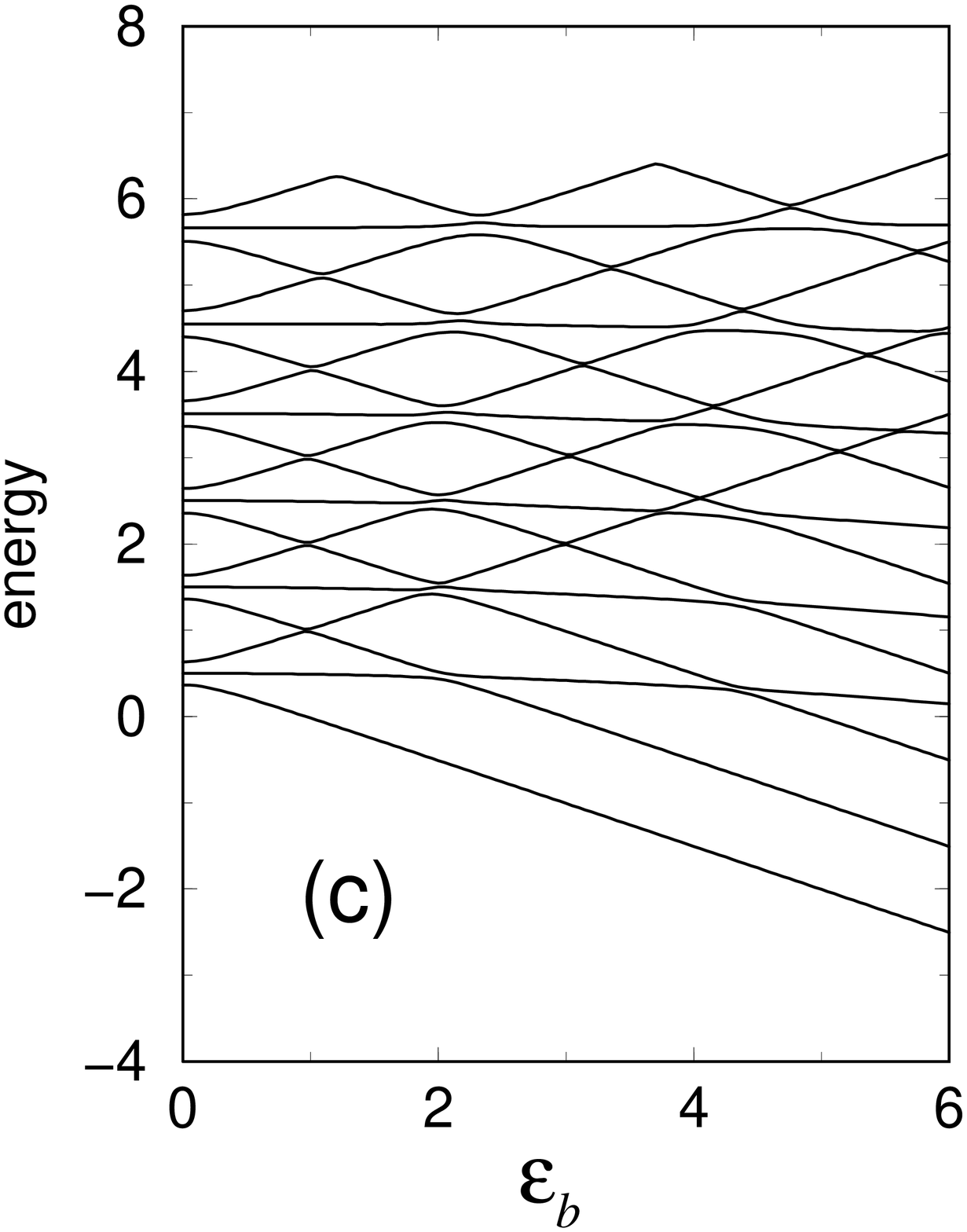, width=6.0cm}
\epsfig{file=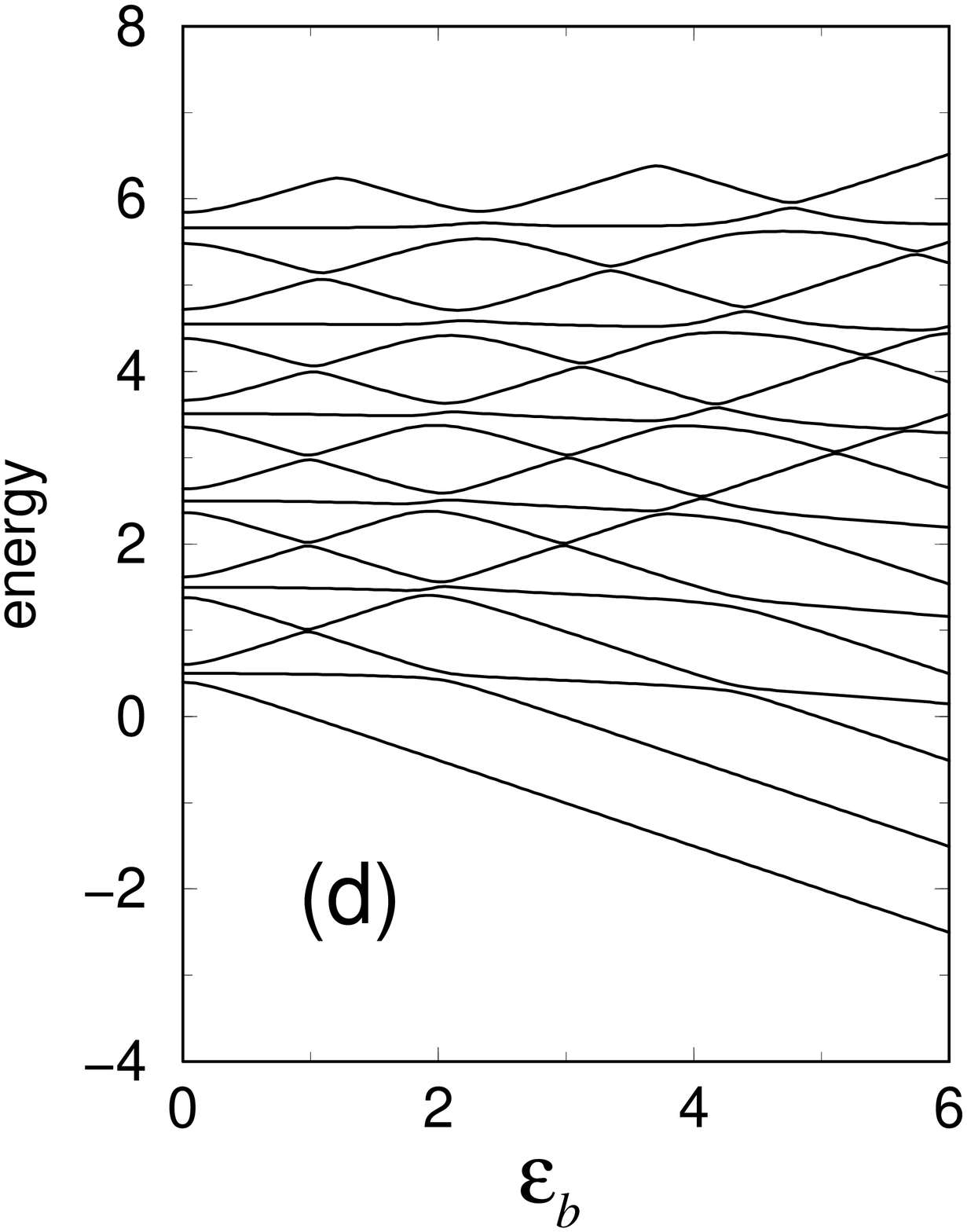, width=6.0cm}}
\caption{\small{Dynamics of the lowest 18 eigenvalues for  systems with
 various parameter values. In the upper panels,
 $\alpha=0.2$, $x_0=5$  with $V=0$ (a)
 and $V=0.5$ (b); in the lower panels
 $V=0.25$, $\alpha=0.2$ (c)
 and $V=0.5$, $\alpha=0.4$ (d).}}
\label{fig:two}
\end{figure}

\begin{figure}[t]
\center{ \epsfig{file=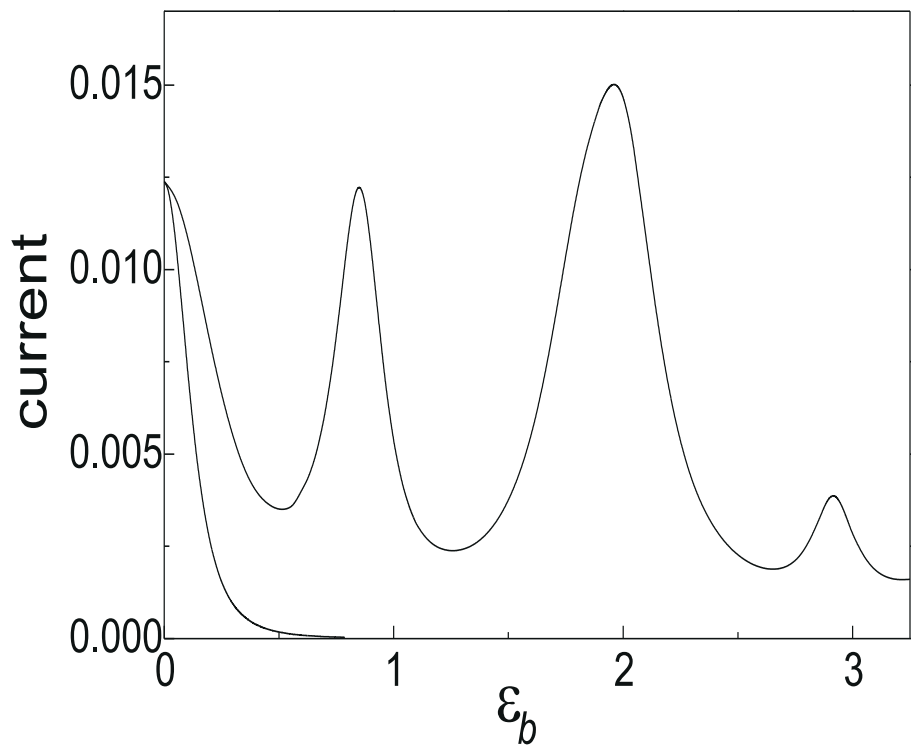, width=12.0cm}}
\caption{\small{Steady-state current through the shuttle with
(upper curve) and without (lower curve) the oscillator; $ V=0.5$,
$\Gamma=0.05$,  $\gamma=0.025$, $x_0=5$ and $\alpha=0.2$.}}
\label{fig:three}
\end{figure}

\begin{figure}
\center{ \epsfig{file=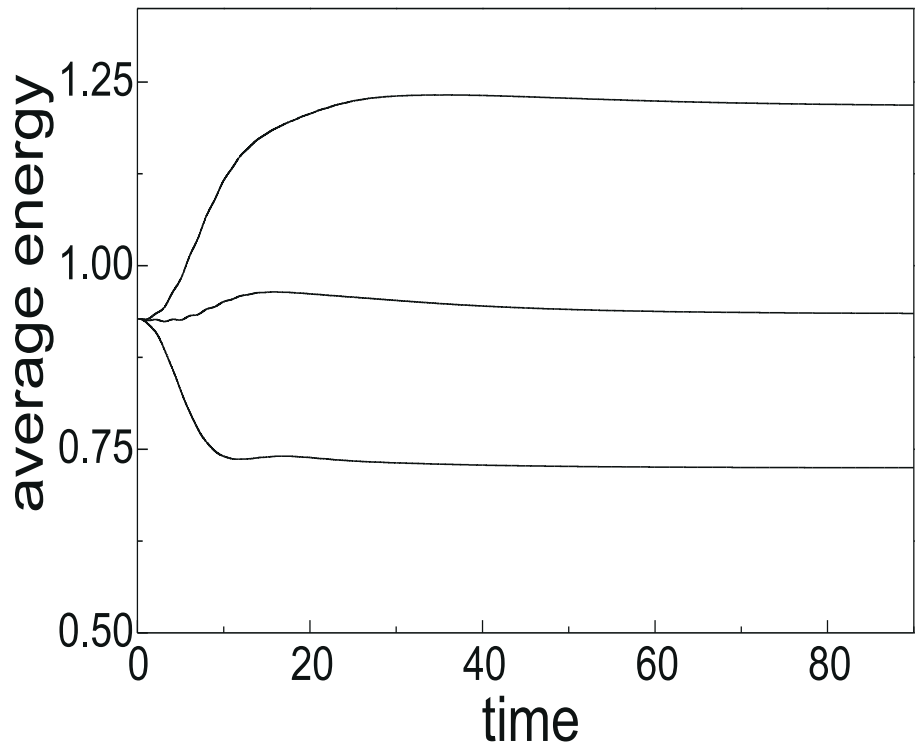, width=12.0cm}}
\caption{\small{Evolution of the energy expectation value of the
shuttle over time (in units of the oscillator period) as a
function of the damping, $\gamma$. From top to bottom, the curves
correspond to $\gamma=0.015,0.025$ and $0.05$, with
$\varepsilon_b=0.86$, $V=0.5$, $\alpha=0.2$, $x_0=5$ and
$\Gamma=0.05$.}} \label{fig:threex}
\end{figure}

\begin{figure}
\center{ \epsfig{file=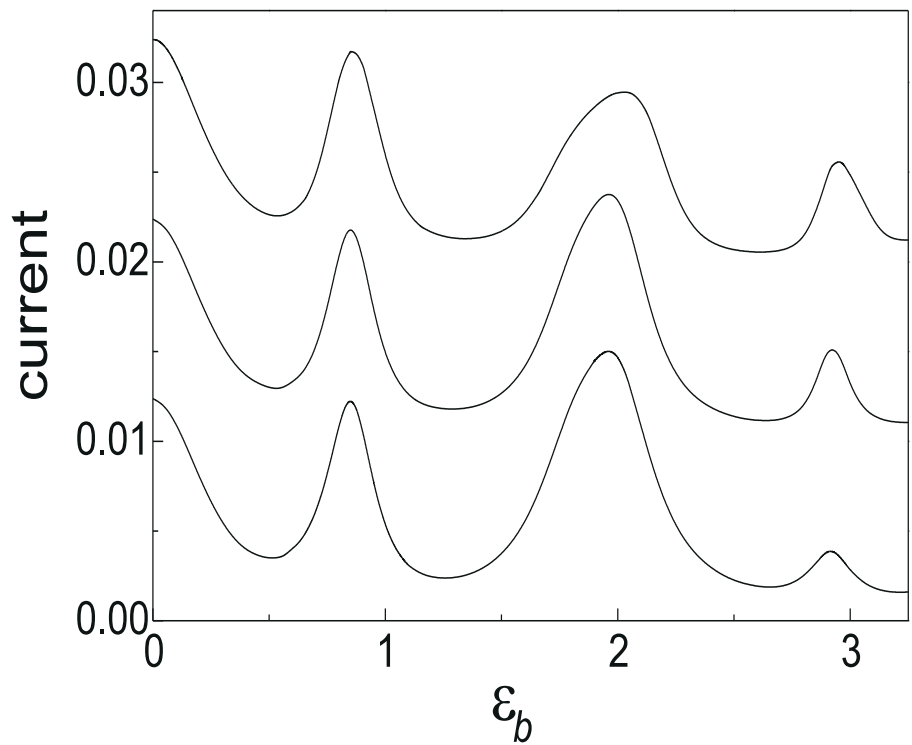, width=12.0cm}}
\caption{\small{Effect of the oscillator damping on the current.
The curves have been displaced for clarity, from top to bottom
they correspond to $\gamma=0.005,0.025$ and $0.05$, with $V=0.5$,
$\alpha=0.2$, $x_0=5$ and $\Gamma=0.05$.}} \label{fig:fiver}
\end{figure}

\begin{figure}
\center{ \epsfig{file=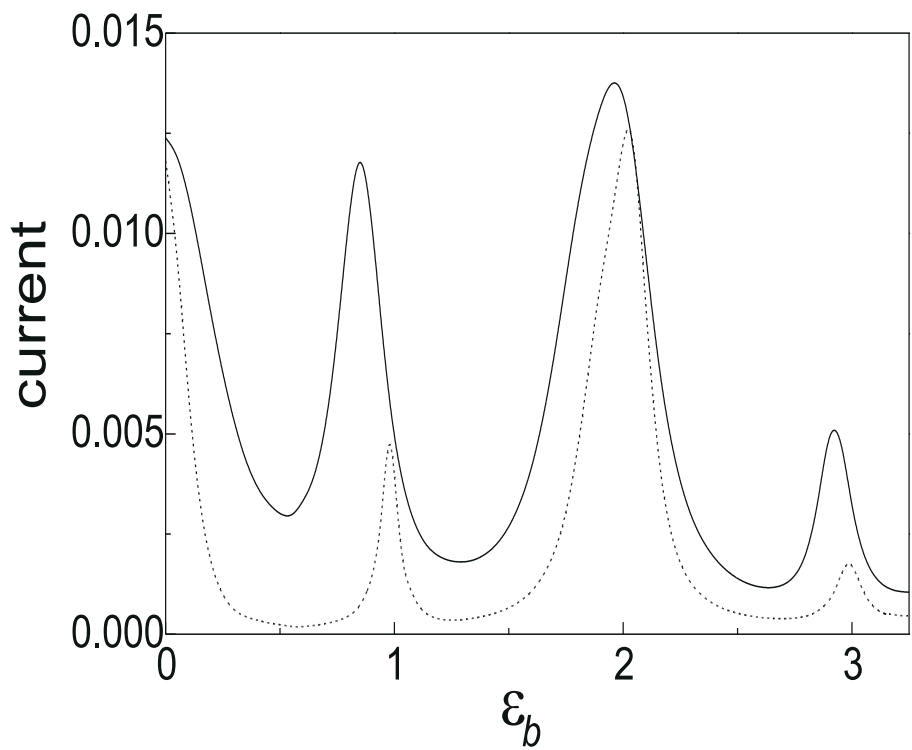, width=12.0cm}}
\caption{\small{Steady state current through the shuttle where $
V=0.5$, $\Gamma=0.05$,  $\gamma=0.025$, $x_0=5$ and $\alpha=0.2$
(continuous line) $0.4$ (dashed line).}} \label{fig:alpha}
\end{figure}

\begin{figure}
\center{ \epsfig{file=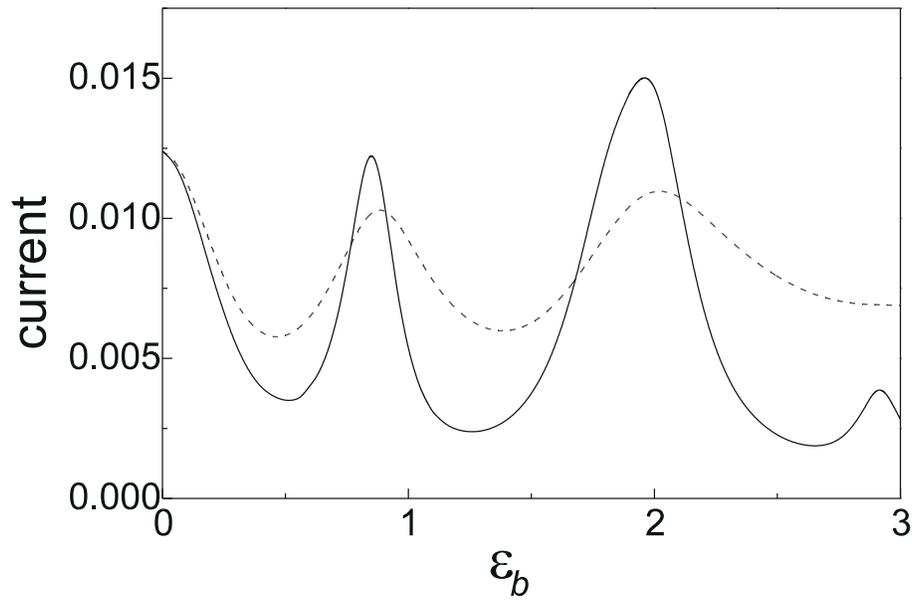, width=12.0cm}}
\caption{\small{Steady state current through the shuttle at
$k_{\rm B}T=3$ (dashed line), compared with that at $k_{\rm B}T=0$
(full line). The parameters are as follows: $V=0.5$ , $\alpha=0.2$
, $\gamma=0.05$, $x_0=5$ and $\Gamma=0.05$.}} \label{fig:four}
\end{figure}

\begin{thebibliography}{99}
\bibitem{comment} The energy level in the central
dot will in fact shift by a fraction of this amount which depend on the details of
the inter-dot capacitances. However, we have neglected this detail
in the interests of simplicity.
\bibitem{note3} The level spacing in the dots also provides a fundamental
limit on the excitation of the oscillator. When the excess energy in the
oscillator approaches the
electron spacing of the electrons in the dots it will be able to return
energy to the electronic degrees of freedom. However, for simplicity we assume
here that the oscillator energy is constrained entirely by damping.
\bibitem{sym} We refer to  the eigenstates as symmetrical in the sense that the probabilities
of finding the electron on the left-hand and right-hand dots are
equal.
\end{thebibliography}
\end{document}